# Acoustic spin-Chern insulator induced by synthetic spin-orbit coupling with spin conservation breaking

Weiyin Deng[1*], Xueqin Huang[1*], Jiuyang Lu[1], Valerio Peri[3], Feng Li[1†], Sebastian D. Huber[3], Zhengyou Liu[2,4†]

[1]School of Physics and Optoelectronics, South China University of Technology, Guangzhou, Guangdong 510640, China

[2]Key Laboratory of Artificial Micro- and Nanostructures of Ministry of Education and School of Physics and Technology, Wuhan University, Wuhan 430072, China

[3]Institute for Theoretical Physics, ETH Zurich, Zürich 8093, Switzerland

[4]Institute for Advanced Studies, Wuhan University, Wuhan 430072, China

*These authors contributed equally to this work.

†Corresponding author. Email: phlifeng@scut.edu.cn; zyliu@whu.edu.cn

**Topologically protected surface modes of classical waves hold the promise to enable a variety of applications ranging from robust transport of energy to reliable information processing networks. The integer quantum Hall effect has delivered on that promise in the electronic realm through high-precision metrology devices. However, both the route of implementing an analogue of the quantum Hall effect as well as the quantum spin Hall effect are obstructed for acoustics by the requirement of a magnetic field, or the presence of fermionic quantum statistics, respectively. Here, we use a two-dimensional acoustic crystal with two layers to mimic spin-orbit coupling, a crucial ingredient of topological insulators. In particular, our setup allows us to free ourselves of symmetry constraints as we rely on the concept of a non-vanishing "spin" Chern number. We experimentally characterize the emerging boundary states which we show to be gapless and helical. Moreover, in an H-shaped device we demonstrate how the transport path can be selected by tuning the geometry, enabling the construction of complex networks.**

The discovery of topological insulators (TIs), featuring a bulk gap and gapless boundary states, opened new avenues for condensed-matter physics [1, 2]. In two spatial dimensions TIs come in two different classes, either described by a Z or $Z_2$ topological index. The first [3-5] break time reversal symmetry (TRS) and are commonly called Chern insulators (CIs). They host an (anomalous) integer quantum Hall effect and their surfaces are characterized by chiral, i.e., unidirectional, surface states. The latter $Z_2$ insulators [6-8], such as the quantum spin Hall effect, are characterized by a pair of gapless helical boundary states. In the presence of spin conservation, the $Z_2$ insulators can equivalently be described by spin-Chern numbers, where the spin sectors might carry an opposite but non-zero Chern number [9]. In fact, spin-Chern numbers are well defined even in the absence of spin conservation or for time reversal symmetry broken systems [9-11]. The spin-Chern numbers have been employed to identify TRS-broken spin-1/2 electronic TIs and pseudospin TIs, giving rise to the concept of spin-Chern insulators (SCIs) [10-17]. The SCIs feature helical boundary states, but whether gapless or not, depends on the system symmetry and micro-structure of the sample boundary [17].

Recently, intense efforts have been devoted to realizing classical analogs of TIs for electromagnetic, mechanical and acoustic waves [18-21]. Photonic CIs have been realized in magneto-optic systems [20-23], mechanical CIs have been proposed in gyroscopic metamaterials [24, 25], acoustic CIs, finally, have been achieved in systems with circulating fluid [26-30]. However, the usage of moving media for acoustics makes these ideas difficult to implement. In the domain of $Z_2$ insulators, Hafezi *et al.* obtained a photonic SCI and observed the helical boundary states in a silicon photonic crystal. The pseudospin-orbit coupling in this system was introduced by the differential optical paths based on ring resonators [31, 32]. In Ref. [33] a mechanical SCI was realized by creating a pseudospin-dependent gauge flux. Both Refs. [32] and [33] rely on the presence of couplings of opposite signs in the equations describing the system dynamics. Such couplings are not natural for low frequency acoustic systems. Nevertheless, helical edge states have been observed also in the acoustic domain. In all these realizations, the topological states have been reported at domain walls along material interfaces [34-43]. This stems from the fact that the pseudospin degrees of freedom in these systems have been realized based on crystalline symmetries that need to be preserved, in particular, at the location of the expected modes. Moreover, the pseudospin is conserved in the aforementioned systems [31-33, 44-46]. This means that these SCIs can actually be viewed as two independent CIs and the topological properties can be described by Chern numbers. When the pseudospin conservation is destroyed, the description in terms of Chern numbers is no longer valid. One then needs to rely on the spin-Chern number to characterize the SCI phases. However, this general case has not yet been explored experimentally.

In this work, we realize an acoustic spin-Chern insulator (ASCI) in a bilayer phononic crystal (PC). By introducing a layer pseudospin degree of freedom, the interlayer coupling induces a synthetic spin-orbit interaction and breaks the pseudospin conservation. Below, we introduce a tight-binding model on a bilayer Lieb lattice hosting the physics of an ASCI. We then map this discrete model to a PC and demonstrate the helical properties, i.e., spin-momentum locking, of the robust boundary states in the ASCI. In addition, we present a switch of helical boundary states in an H-shaped device that holds promise for technological applications.

To illustrate how to realize an ASCI, we construct a tight-binding model on a bilayer Lieb lattice with a unit cell containing three sites in each layer, denoted A (red sphere), B (green sphere), and C (blue sphere) in Fig. 1a. The Hamiltonian is

$$H = t_0 \sum_{\langle ij \rangle, \alpha} c_{i\alpha}^\dagger c_{j\alpha} + m \sum_{i \in A, \alpha} c_{i\alpha}^\dagger c_{i\alpha} + \lambda \sum_{\langle\langle ij \rangle\rangle, \alpha \neq \beta} v_{ij,\alpha} c_{i\alpha}^\dagger c_{j\beta}, \qquad (1)$$

where $c_{i\alpha}^\dagger$ is the creation operator of layer pseudospin $\alpha$ on site $i$. The first term describes the nearest-neighbor intralayer hopping with strength $t_0$. The second term denotes the on-site energy $m$ on site A. The last term represents the chiral interlayer coupling with strength $\lambda$, where $v_{ij,\alpha} = \left[\varepsilon_\alpha (\hat{e}_{kj} \times \hat{e}_{ik})_z + 1\right]/2$ with $\varepsilon_{\uparrow\downarrow} = \pm 1$, $i$ and $j$ are two next-nearest-neighbor sites with $i \neq j$, $k$ is their unique common nearest-neighbor site and the unit vector $\hat{e}_{kj}$ points from $j$ to $k$ (see Supplementary Material I for details). The band dispersion of the model is presented in Fig. 1b. The interlayer coupling opens two bulk gaps and can induce topological phase transitions in the model.

The topological properties of this system can be captured by a spin-Chern number. One can introduce a pseudospin $\tau_\alpha = \sigma_\alpha \otimes I_3$, where the Pauli matrices $\sigma_\alpha$ act on the layer degree of freedom. While none of the components of $\tau$ is conserved, one can use the projection of, say $\sigma_y$ into pairs of bands below the gap to split them. These split bands lead to well-defined fiber bundles which may carry a non-zero Chern number: the spin-Chern number. These spin-Chern numbers are a tool well tailored to classical systems, as they neither require any symmetry nor the presence of a fermionic time reversal operator. However, it is important to note, that one relies not only on a spectral gap, but also on the spin-projection gap that allows for the splitting of the bands. Moreover, the details of the edge physics has to be inspected independently of the bulk, as there might be a spin-gap closing induced by the surface termination [10, 11], see Supplementary Material II for details.

For illustrations, we focus on the topological properties of the lower gap. Figure 1c shows the spin-Chern number of the lower two bands $C_s^l$ as a function of $\lambda/t_0$ and $m/t_0$. Three topologically distinct phases exist. At the phase boundaries, indicated by the white lines, the energetic bulk gap closes. In the absence of spin-orbit interaction ($\lambda = 0$), the mass term opens a trivial gap: $C_s^l = 0$. The projected band dispersion for a ribbon with $C_s^l = 2$ is plotted in Fig. 1d. The boundary states of the two lines oriented along the dotted black arrows localize at one boundary of the ribbon (the inset), while the others localize at the other boundary. The complete topological phases and associated helical boundary states are shown in Supplementary Material II and III, respectively.

We now consider a PC implementation of the SCI for acoustic waves. As shown in Fig. 2a, the PC sample, fabricated by 3D printing, consists of a bilayer structure with interlayer connections realized by chiral tubes. Each layer of the unit cell contains three nonequivalent cavities connected by intralayer tubes (Fig. 2b). Mapping the PC to a tight-binding model, the cavities can be regarded as lattice sites, while the tubes provide hopping terms. The square unit cell has in-plane length $a = 20$ mm and height $h =$ 12.5 mm. The three square cavities composing a layer of the unit cell have the same height $h_c = 5$ mm and different in-plane dimensions: $L_A = 7$ mm and $L_B = L_C = L_0 = 8$ mm. The width and height of the intralayer tubes are $L_t = 3.2$ mm and $h_t = 3$ mm, respectively. The diameter of the interlayer tubes is $d = 3.2$ mm. Since the volume of cavity A is smaller than those of the other cavities, modes localized there are detuned to higher frequencies. This corresponds to a regime $m/t_0 < 0$ for the tight-binding model of Eq. (1): a region of the phase diagram with $C_s^l = 2$.

In Fig. 2c, we present the measured bulk band dispersion along high symmetry lines. Overlaid to the experimental data, we show the simulated bands. A bulk gap at M opens thank to the chiral interlayer couplers. To confirm the topologically non-trivial nature of the gap, we calculate the spin-dependent Berry curvature and spin-Chern number of the lowest two bands in the real PC. As the pressure field is mainly localized at the cavities, we construct the normalized wavefunctions $\varphi(\mathbf{k})$ of the PC by using the pressure field sampled at the center of each cavity. Using these wavefunctions (Supplementary Material II), we obtain the spin-dependent Berry curvatures $\Omega_\pm^l(\mathbf{k})$, shown in Fig. 2d. The two spin-projection sectors have opposite Berry curvature. By integrating separately $\Omega_\pm^l(\mathbf{k})$ in the whole Brillouin zone, we determine the spin-Chern number of the PC. The result, $C_s^l = 2$, confirms that the PC has a gap with the same non-trivial topology predicted by the tight-binding model. The other phases of the ASCI, corresponding to the cases of $\lambda/t_0 > 0$, are studied in Supplementary Materials IV.

The nonzero $C_s^l$ can induce a pair of helical boundary modes. The projected band dispersions along the $k_x$ direction are plotted in Figs. 3a and 3b for the whole-cell and half-cell boundaries, respectively. The color maps represent the experimental dispersions, while the overlaid lines are the result of full wave simulations. A pair of counterpropagating gapless boundary states (solid white lines) exists in the gap for both boundaries. The spin polarization along the $y$ direction is defined as $\langle \tau_y \rangle = \langle \varphi_{k_x} | \tau_y | \varphi_{k_x} \rangle$, where $\varphi_{k_x}$ is the eigenmode of the projected dispersion of the PC ribbon sampled at the center of each cavity. The $x$ and $z$ components of the spin have little contribution. The polarizations $\langle \tau_y \rangle$ of boundary states for $k_x < 0$ are opposite to those for $k_x > 0$, as shown in Figs. 3c and 3d where lines represent simulated values and circles experimental ones. The pair of boundary states on the same boundary are helical, as expected for a SCI. The experimental results are in excellent agreement with the simulations.

The helical nature of the surface states renders them ideal for the realization of a spin-filtered one-way waveguide. Figure 4a shows the transport of acoustic boundary wave at 7.44 kHz in a sample possessing a rectangular defect. The boundary waves propagate smoothly around the defect. Good agreement is found between the experimental (upper panel) and simulation results (lower panel), although the measured boundary waves attenuate during propagation due to air damping. In Fig. 4b, we present the ratio of the energy flows between an output and input channel for a path with the rectangular defect and a straight one of the same length. The transmission of the two samples agree in the bulk gap, indicating the robustness of the surface modes against backscattering induced by the defect.

In the electronic system, a switch effect of the topological boundary states by means of a quantum point contact has been predicted [47]. Here, we experimentally realize such a novel transport phenomenon for the ASCI boundary waves. Figure 5a (top panel) shows a schematic of the H-shaped structure of the ASCI, where the width of the left and right ribbons is $W = 20a$, and the length of the middle ribbon ("bridge") is $L_\mathrm{m} = 13a$. Due to the spin-momentum locking, the boundary waves excited from channel 1 can only propagate to terminals 2 and 3, while they cannot be detected at terminal 4 (see Fig. 5a). In the middle panel of Fig. 5a, we show the lowest (highest) frequency $f_\mathrm{b}^l$ ($f_\mathrm{b}^h$) of the upstream (downstream) branch of the boundary states in the lower gap as a function of the width of the middle ribbon $W_\mathrm{m}$ (for more details, see Supplementary Material V). When $W_\mathrm{m}$ is sufficiently large, as shown in Fig. 3, the boundary wave dispersions on the opposite boundaries are gapless due to the lack of interactions between them. Hence, $f_\mathrm{b}^l = f_\mathrm{b}^h$. By gradually decreasing $W_\mathrm{m}$, the overlap of the surface waves localized at opposite boundaries can open a gap in the dispersions of the edge states. This results in a frequency mismatch between $f_b^l$ and $f_b^h$. We calculate the transmissions from channel 1 to terminals 2 and 3 at 7.44 kHz and show the results in the bottom panel of Fig. 5a The boundary waves can cross the bridge and reach terminal 3 for $W_m > 4a$ and are completely blocked from terminal 2 for $W_m < 2a$. The crossover from one to zero for $S_{31}$ (transmission between channel 1 and terminal 3) occurs sharply as a function of $W_m$, indicating ideal partition transport of the boundary waves in the H-shaped ASCI.

In Figs. 5b-5d, the pressure distributions of the boundary waves are shown for three different $W_m$ values from wide to narrow. The upper (lower) panel is the measured (calculated) result. For $W_m = 5.5a$, the boundary states completely propagate to terminal 3 because of the gapless boundary wave dispersions. For $W_m = 2.5a$, the boundary waves can partly propagate both to the right and back to the left ribbon. For $W_m = 1.5a$, the boundary waves are mainly blocked to terminal 2 since the finite size effect opens a large gap in the boundary wave dispersions. Due to air losses, the measured transmission of the boundary waves cannot reach unity, as in the simulated results in the absence of dissipations. However, the partition behavior shows good agreement between the simulations and the experiment. These results provide an efficient way to control the boundary waves.

In summary, we have realized an ASCI with a pair of helical boundary states, which is of fundamental interest and opens up an avenue for applications of topological acoustics. Our simple system is ideal for studying the physics of SCIs as it is of macroscopic scale. The topological helical boundary states may have potential applications in innovative acoustic devices, such as topological splitters/switches with high tolerance.

**Data availability**

The data that support the plots within this paper and other findings of this study are available from the corresponding author upon reasonable request.

**Acknowledgements**

This work is supported by the National Natural Science Foundation of China (Grant Nos. 11804101, 11890701, 11572318, 11604102, 11704128, 11774275, 11974120 and 11974005), the National Key R&D Program of China (Grant No. 2018FYA0305800), Guangdong Innovative and Entrepreneurial Research Team Program (Grant No. 2016ZT06C594), and the Fundamental Research Funds for the Central Universities (Grant Nos. 2018MS93, 2019JQ07, and 2019ZD49). VP and SDH acknowledge support from the European Research Council under the Grant Agreement No. 771503.

**Author contributions**

All authors contributed extensively to the work presented in this paper.

**Competing financial interests**

The authors declare no competing financial interests.

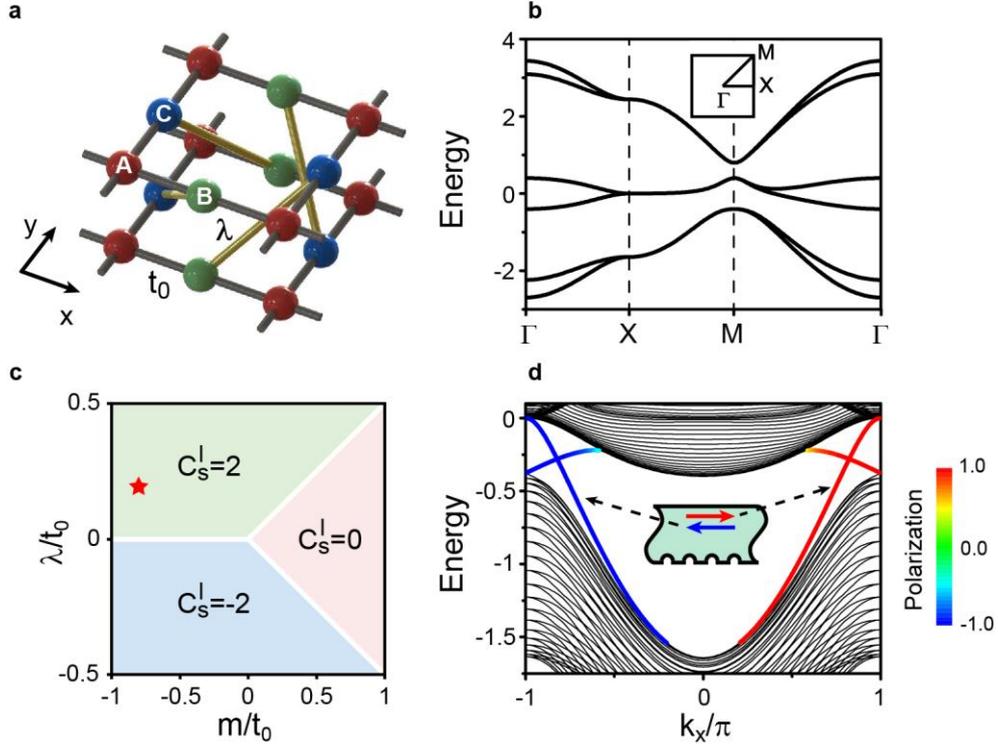

**Figure 1 | SCI and helical boundary states for a bilayer Lieb lattice model. a,** Schematic of the lattice structure. The red, green, and blue spheres of each layer denote A, B, and C lattices. **b,** The bulk band structure along high symmetry lines. The interlayer coupling gives rise to two band gaps. Inset: the first Brillouin zone. **c,** Phase diagram determined by the spin-Chern number of the lower two bands $C_s^l$ in the $\lambda/t_0$ and $m/t_0$ plane. The white lines represent lower band gap closure. The red star denotes the phase with the specific parameters used in **b** and **d**. **d,** The boundary state dispersion of a ribbon with $C_s^l = 2$ in the lower gap. A pair of boundary states at the same edge (inset) have opposite layer pseudospin polarizations (red and blue colors). The parameters are chosen as $t_0 = -1$, $\lambda = -0.2$, and $m = 0.8$.

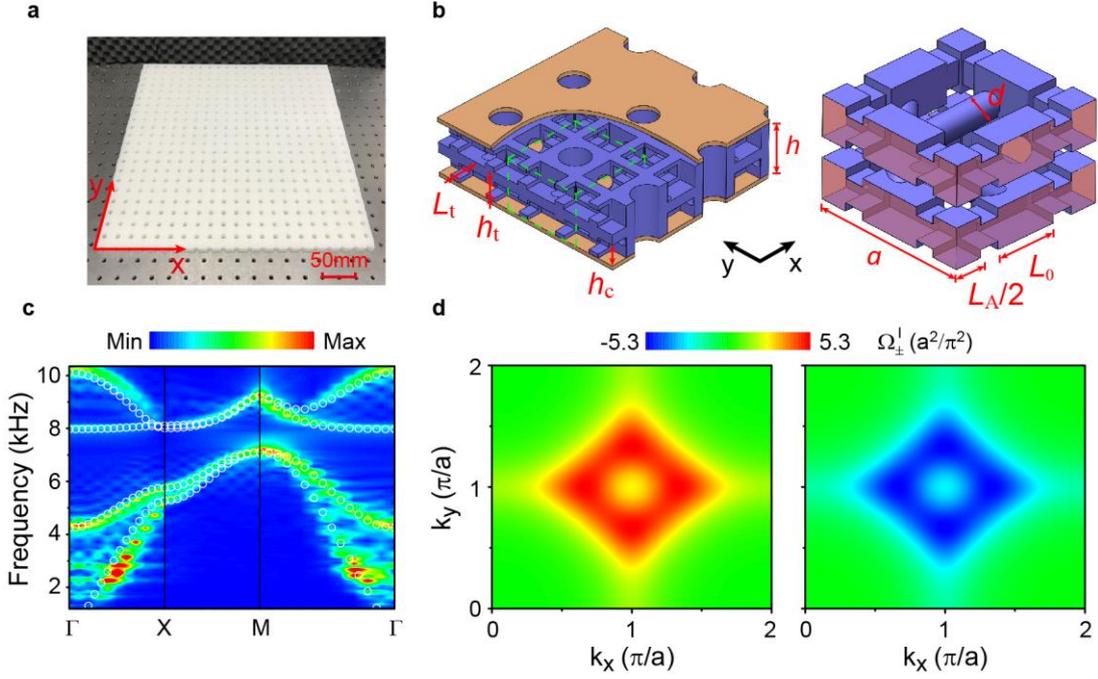

**Figure 2 | ASCI and bulk properties. a,** A photo of the bilayer sample. The cylindrical holes are designed for cost savings and have no effect on the propagation of acoustic waves in the PC. **b,** Left panel: magnified side view of the sample, which contains the whole-cell (up edge) and half-cell (down edge) boundaries along the *x* direction. Air fills the inside of the bilayer structure (blue color) between two rigid plates (brown color). Right panel: the unit cell of the sample, corresponding to the green dotted box in the left panel. The blue (pink) areas represent the rigid (periodic) boundaries. **c,** The bulk band structure of the lowest four modes along high symmetry lines. The color maps denote the measured data, and the white circles represent the simulated results. **d,** The calculated Berry curvatures of the spin up (left) and down (right) projection sectors for the lower two bands.

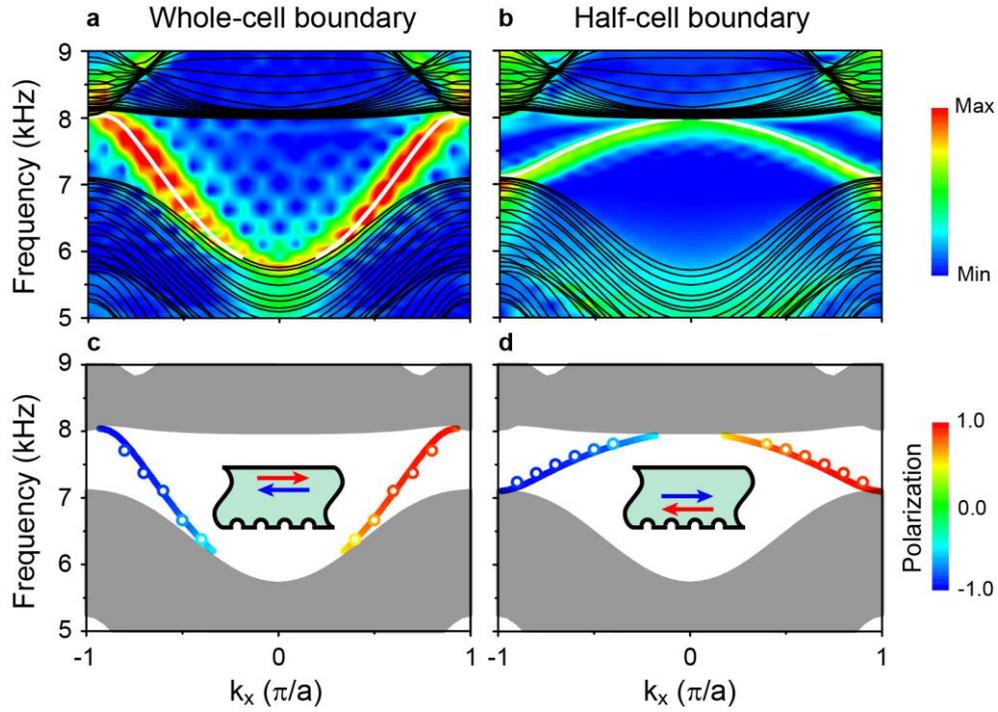

**Figure 3 | Acoustic helical boundary waves. a**, **b,** The dispersions of helical boundary waves on the whole-cell and half-cell boundaries, respectively. The color maps denote the measured data, and the white and black lines represent the simulated dispersions of the boundary states and projected bulk states, respectively. **c**, **d**, The spin polarizations of the boundary waves for the whole-cell and half-cell boundaries, respectively (lines for simulations and circles for experimental results). Inset: the red (blue) color of arrows denotes spin up (down). A pair of gapless boundary waves with opposite spin polarizations counter-propagate along the boundary.

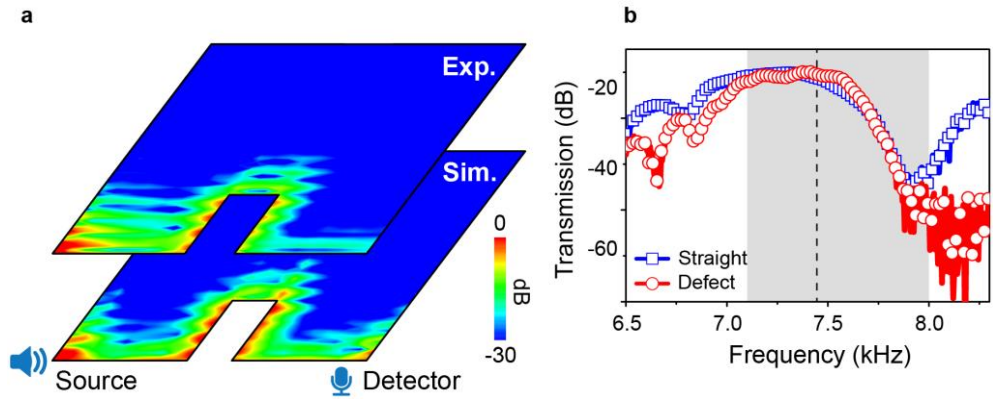

**Figure 4 | Experimental demonstration of backscattering robustness of the acoustic boundary waves for the whole-cell boundary. a,** The field distributions 7.44 kHz (dashed line in **b**) from the simulation and experiment for a sample possessing a rectangular defect. The loudspeaker and microphone symbols denote the source and detector positions, respectively. **b,** The measured transmission for the defect path, compared with that for a sample having a straight path with the same length. The good agreement in the bulk gap (shadowed region) indicates the negligibly weak backscattering of the acoustic boundary waves.

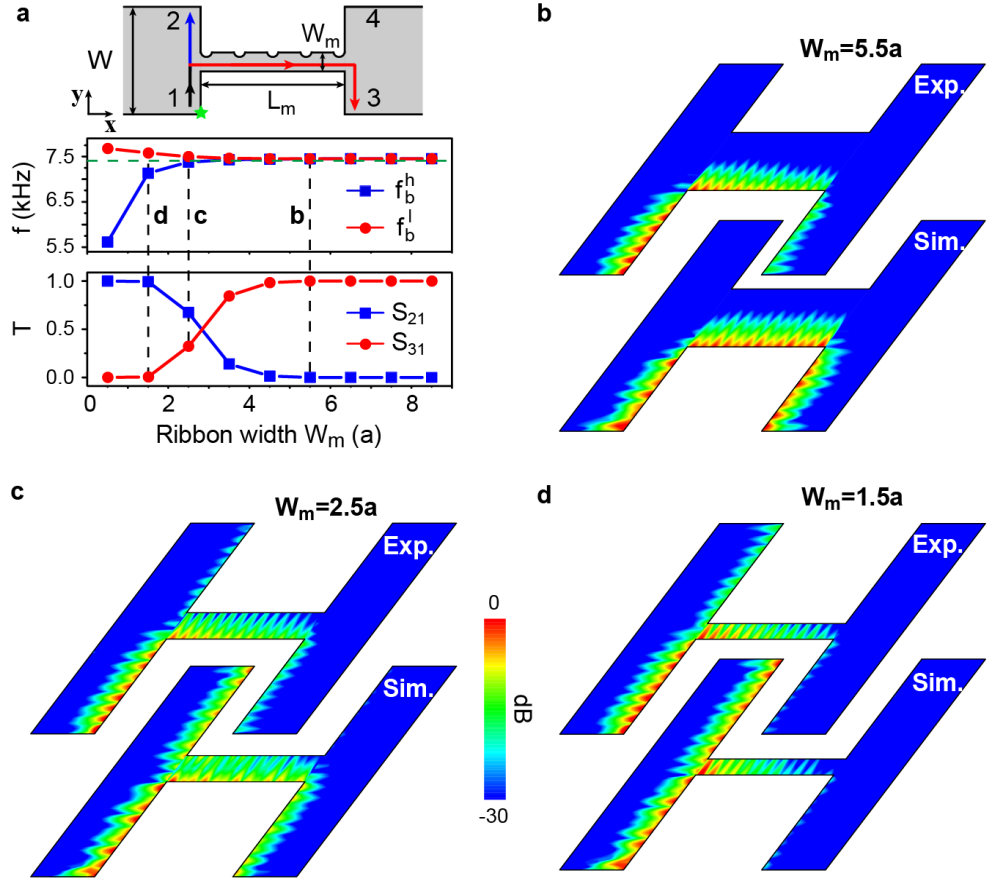

**Figure 5 | The partition behavior of the acoustic boundary waves in an H-shaped ASCI. a,** Top panel: the schematic of the H-shaped PC, where the green star denotes the position of the source. Middle panel: the lowest (highest) frequency $f_b^l$ ($f_b^h$) of the upstream (downstream) branch of the boundary states in the lower gap as a function of the width of the middle ribbon $W_m$. The dashed green line denotes the excitation frequency of 7.44 kHz. Bottom panel: the calculated transmission from channel 1 to terminals 2 ($S_{21}$) and 3 ($S_{31}$). **b-d,** The simulated and experimental field distributions in the H-shaped PC for three different $W_m$ (the dashed black lines in **a**). The boundary waves can be controlled by tuning $W_m$.